# Standardization of Proton Induced X-Ray Emission for Analysis of Trace Elements in Thick Targets

Johar Zeb[1,2], Shad Ali[1,2], Muhammad Haneef[1], Azhar Muhammad Naeem[3], Jehan Akbar[1,4]

[1]Department of Physics Hazara University Mansehra, [2]Department of Physics, Beijing Normal University, China

[3]Department of Electrical Engineering, University of the Punjab, Lahore, Pakistan, [4]International Center for Theoretical Physics, Italy

E-mail: Jehan@ictp.it

This paper presents the standardization of Proton Induced X-rays Emission (PIXE) technique for the trace element analysis of thick standard samples. Three standard reference materials (SRMs) viz-à-vis titanium, copper and iron base alloys were used for the study due to their availability. The protons beam was accelerated up to 2.57 MeV energy by 5UDH-II tandem Pelletron accelerator and samples were irradiated at different geometry and durations. Spectrum was acquired using a multi-channel spectrum analyzer while spectrum analysis was done using a GUPIXWIN model for determination of elemental concentrations of trace elements. The obtained experimental data was compared with theoretical data and results were found in close agreement.

## 1. INTRODUCTION

The word PIXE is an acronym which stands for Particle/Proton Induced X-rays Emission. PIXE was first experimentally shown by Sven A.E. Johansson of Lund Institute of Science and Technology in 1970 [1]. This technique is very useful due to its non-destructive nature and multi-elemental analysis simultaneously. Due to its longer probe depth (tens of μm.) the PIXE can be performed for a large range of elements ($Na \leq Z \leq U$) of the periodic table. PIXE is fast, highly sensitive (ppm), precise, accurate and is better for analysis of both matrix and trace elements in thick samples. In addition, PIXE give you bremsstrahlung radiation, high fluorescence yield and X-ray production cross section [1]. Other ion beam techniques such as PIGE is better than PIXE when measuring some elements especially the low Z elements as self-absorption effect is completely avoided [2, 3]. PIXE has many applications in the study of archaeology, environmental, biological, medicinal and forensic materials [4-9].

However, despite the numerous advantages of PIXE technique, standardization or calibration with certified reference materials before each experiments poses a major challenge to most researchers [10]. We report on the standardization of PIXE for accurate analysis of trace elements in thick samples

## 2. EXPERIMENTAL
### 2.1 PIXE Set Up

The experimental setup used for PIXE analysis is shown in Figure 1. For inducing X-rays, a high energy proton beam (2.57 MeV) accelerated by 5UDH-II tandem Pelletron accelerator excites the atoms of target material. Titanium based alloy, copper based alloy and iron based alloy were used as target materials for the analysis of trace elements. The excited atoms on its de-excitation emit characteristic X-rays in the form of K, L and M X-rays spectrum [3]. The emitted X-rays are passed through a thin window (Mylar foil 100 μm thick made from ultra-thin polymer) and detected using a Silicon Drift Detector (SDD) placed at an angle of $45^0$ to the target for good resolution (about 138 eV at 11 KeV energy. The detected signal is amplified and transferred to the readout. Faraday cup was placed to collect the beam charge and a current integrator records the total collected beam charge during a run. The whole experiment was done under vacuum chamber at a pressure of about $10^{-6}$ torr to reduce the background radiation. After the PIXE spectrum was obtained, GUPIXWIN software was used for quantitative analysis of the data [12]. The use of GUPIXWIN software is preferred to other software programs due to its good status of databases (e.g. cross-sections, fluorescence yield, Coster–Kronig probabilities, stopping powers and attenuation coefficients etc.) and more elemental fitting [13].

### 2.2 SPECTRA ACQUISITION

For quantitative analysis, PIXE spectrum of each SRM was obtained on different dates for three different values of incident beam charge. For PIXE calibration, standard reference materials (National Institute of science and technology) were irradiated, analyzed and the results were compared with theoretical values for two samples, NIST654B titanium and copper alloy 99 % pure, nine data spectra were obtained by varying the beam charge (0.5 μC, 1μC and 1.5μC) on three consecutive working days for NIST SRM 1155 (Cr 18-Ni 12-Mo 2). Incident beam energy 2.57 MeV and current beam 2nA are used for irradiation of the sample and Silicon drift detector is placed at an angle of $45^0$ to the target to detect the signal. The range of energy and beam charges is selected in order to increase the fluorescence yield from the sample and decrease bremsstrahlung radiation [11].

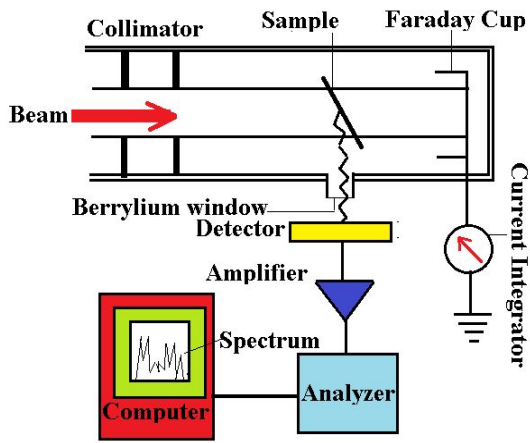

Figure 1: Schematic of PIXE experimental set up.

## 2.3 GUPIXWIN Analysis

After obtaining the spectrum files, the next important step to get required accuracy is the quantitative analysis by using GUPIXWIN software. Using GUPIXWIN software with suitable parameters, we can determine both elemental composition and concentration present in the target. In GUPIXWIN software "Fixed Matrix solution" method is taken to determine the trace elements in the certified sample. In our experiment, the incident beam energy 2.57 MeV and current beam 2nA are used. Position of detector, target and the angle made by detector with target must be fixed during experiment, we only need to determine the instrumental constant (H-value) against the energy data file, i.e. HED file. Finding the value of instrumental constant "H" is also very important for proper calibration of the PIXE technique.

## 2.4. Instrumental Constant (H-Value) Verses Energy Data (HED) File

The H quantity is a standardization factor that depends on elements, which translates the peak area of elements of the collected charge into concentration of elements. Initially, in an ideal case, H is taken equal to the solid angle (Ω), that made by the detector in the geometry of the chamber. For our experimental system, H-value is equal to 0.00153 (using formula $\Omega = H = \frac{A}{d^2}$).

Practically it slightly depends on the elements atomic number (Z), which produces inaccuracies in detector description and also in data bases [14].

Table 1(a): Combined K HED file.

| K X-Rays File | |
|---|---|
| Energy(KeV) | H-Value |
| 1.487 | 2.5643 |
| 1.74 | 1.95 |
| 2.014 | 0.0145 |
| 2.013 | 0.0077 |
| 4.51 | 0.001147 |
| 4.952 | 0.000975 |
| 5.415 | 0.00096 |
| 6.403 | 0.00098 |
| 6.93 | 0.00415 |
| 7.478 | 0.00097 |
| 8.047 | 0.001288 |
| 8.905 | 0.001118 |
| 10.54 | 0.0049 |

Table 1(b): Combined L HED file.

| L X-Rays File | |
|---|---|
| Energy(KeV) | H-Value |
| 1.8066 | 0.53 |
| 2.2932 | 0.00099 |
| 2.3948 | 1.99 |
| 3.444 | 0.00542 |
| 8.3976 | 0.00053 |
| 10.551 | 0.195 |

The standard H-value can be determined, either by constant method or by energy dependence method. If geometrical setup is completely accurate then constant method is used and if there is any error, then energy dependence method is used. In the later method, HED data file is required. The HED file can be created in note-pad or word-pad with a desired name and need to be saved in the example folder. Furthermore, in the HED file a separate file is present for each (K, L and M) X-ray measurements. It is required to write H-value in front of the high energy (having 100% intensity in the keV) X-ray values followed by a tab in separate lines for each file (K, L or M X-ray). We need to change the H-values (either increase or decrease) in these files according to the output data result. We have created HED file in two steps.

First, an individual HED file is created for each analyzed sample. Secondly, these HED files values are combined to form a combined HED file, which is used for the analysis of all samples. The combined HED files are given in the Table 1(a) and (b) respectively. These HED files are used to create the PAR-file for the given samples. The final concentration obtained is compared with SRM's concentration. Both of concentrations were found to be in good agreement with each other.

## 3. RESULTS

All samples are calibrated with copper sample using calibration method as previously discussed. The discussion of each sample in detail is given in the following sub-sections.

### 3.1 NIST654B Titanium Alloy Sample

The PIXE curve for NIST654B titanium alloy sample is shown in the Figure 2.

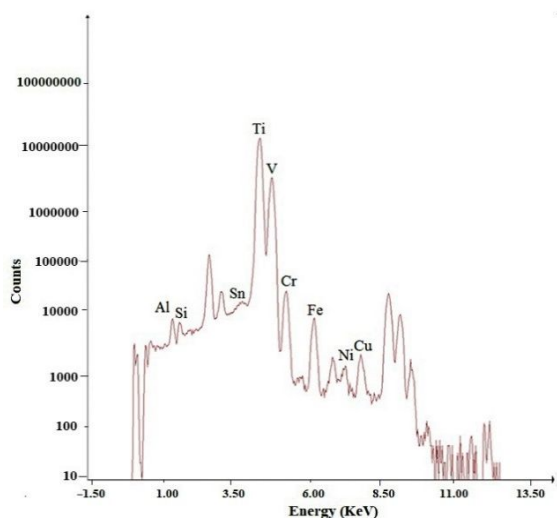

This spectrum shows both the trace and matrix elements composition and concentration present in sample. As shown in the figure, the identified elements in the sample are Ti, V, Al, Fe, Si, Ni, Cr, Cu and Sn. A parameter (PAR) file of this sample is created, which consist of all information about it. When the PAR-file is completed with a proper sequence, run the GUPIX engine to get output data. For NIST654B titanium alloy sample, the concentration of analyzed data, standard data and their standard deviation ($\sigma$) are given in Table 2. The dispersion plot ($\sigma$) between standard and analyzed data is shown in Figure 3. The standard deviation ($\sigma$) values are small, which represent good agreement between measured and already known data.

Figure 2: Measured PIXE spectrum for NIST654B alloy sample.

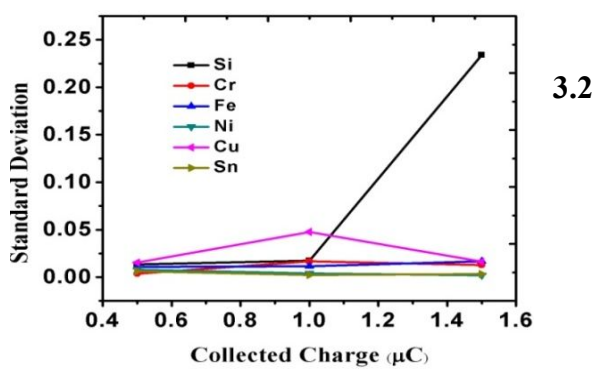

### 3.2

Figure 3: Dispersion plot for NIST654B alloy sample.

### NIST1155 Iron Alloy Sample

The PIXE curve for NIST1155 iron alloy sample is shown in the Figure 4.

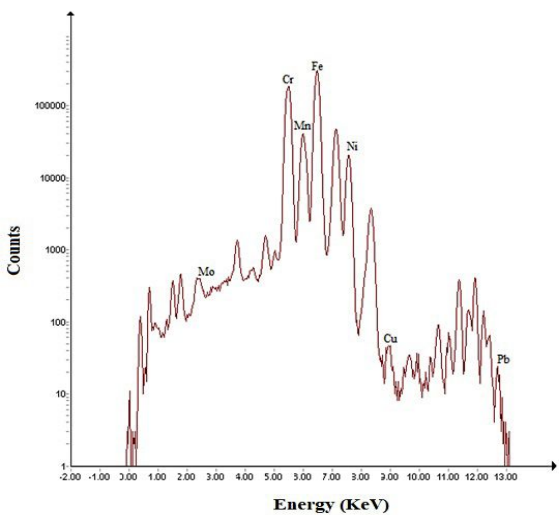

The matrix and trace elements which were identified in

Figure 4: PIXE spectrum for NIST1155 iron alloy sample.

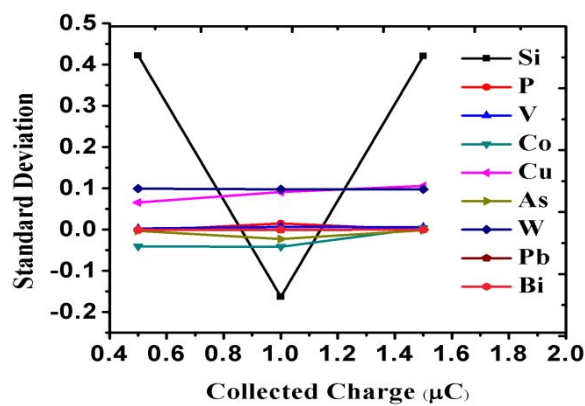

Figure 5: Dispersion plot of NIST1155 iron alloy sample.

this sample are Co, P, Si, V, Cr, Mn, Fe, Ni, Cu, As, Bi

and Pb. A parameter (PAR) file is also created using GUPIXWIN software. After running the GUPIXWIN engine, the final output data is recorded as shown in the Table 3. The data was measured on a single date. Difference between measured data and already known data were shown in Table 3. The dispersion plot for this sample is shown in the Figure 5. The difference between concentration of measured and already known data values is negligibly small.

Table 2: Comparison of measured and actual composition of NIST654B alloy sample for different beam of charges at different dates.

| Charge(μC) | Element | Actual (a) % | $b_1$ % | $b_2$ % | $b_3$ % | <b>% | a-<b>% | Standard Deviation(σ)% |
|---|---|---|---|---|---|---|---|---|
| 1.5 | Si | 0.045 | 0.0465 | 0.5503 | 0.0484 | 0.213 | -0.168 | ±0.234 |
|  | Cr | 0.025 | 0.1508 | 0.1493 | 0.1246 | 0.1415 | -0.1165 | ±0.0128 |
|  | Fe | 0.23 | 0.2818 | 0.2551 | 0.2416 | 0.2595 | -0.0295 | ±0.0167 |
|  | Ni | 0.028 | 0.0298 | 0.027 | 0.0251 | 0.0273 | 0.0007 | ±0.0019 |
|  | Cu | 0.004 | 0.0086 | 0.0482 | 0.0345 | 0.0304 | -0.0264 | ±0.0165 |
|  | Sn | 0.023 | 0.0230 | 0.0258 | 0.0187 | 0.0225 | 0.0005 | ±0.0032 |
| 1 | Si | 0.045 | 0.0474 | 0.0808 | 0.0420 | 0.0569 | 0.0119 | ±0.0174 |
|  | Cr | 0.025 | 0.1850 | 0.1563 | 0.1454 | 0.1621 | -0.1371 | ±0.0166 |
|  | Fe | 0.23 | 0.2753 | 0.2666 | 0.2512 | 0.2643 | -0.0343 | ±0.0115 |
|  | Ni | 0.028 | 0.0326 | 0.0284 | 0.0233 | 0.0281 | -0.0001 | ±0.004 |
|  | Cu | 0.004 | 0.0130 | 0.0424 | 0.0347 | 0.03 | -0.026 | ±0.0477 |
|  | Sn | 0.023 | 0.0202 | 0.0263 | 0.0227 | 0.0231 | -0.0001 | ±0.00246 |
| 0.5 | Si | 0.045 | 0.0555 | 0.0726 | 0.0399 | 0.056 | -0.011 | ±0.0133 |
|  | Cr | 0.025 | 0.1764 | 0.1793 | 0.1790 | 0.1782 | -0.153 | ±0.0036 |
|  | Fe | 0.23 | 0.2616 | 0.2416 | 0.2634 | 0.2555 | -0.0255 | ±0.0107 |
|  | Ni | 0.028 | 0.0386 | 0.0277 | 0.0195 | 0.0286 | -0.0006 | ±0.0078 |
|  | Cu | 0.004 | 0.0091 | 0.0426 | 0.0412 | 0.03096 | -0.027 | ±0.0154 |
|  | Sn | 0.023 | 0.0179 | 0.0291 | 0.0148 | 0.0206 | 0.0024 | ±0.0061 |

Table 3: Comparison of measured and actual composition of NIST1155 iron alloy sample for different beam charges.

| Charge (μC) | Element | Actual (a)% | $b_1$% | (a-b)% |
|---|---|---|---|---|
| 1.5 | Si | 0.5093 | 0.0889 | 0.4204 |
|  | P | 0.0200 | 0.0201 | -0.0001 |
|  | V | 0.050 | 0.0437 | 0.0063 |
|  | Co | 0.109 | 0.1055 | 0.0035 |
|  | Cu | 0.175 | 0.0690 | 0.106 |
|  | As | 0.01067 | 0.0107 | -0.00173 |
|  | W | 0.11 | 0.0124 | 0.0976 |
|  | Pb | 0.001 | 0.0011 | -0.0001 |
|  | Bi | 0.0004 | 0.0002 | 0.0002 |
| 1 | Si | 0.5093 | 0.6723 | -0.163 |
|  | P | 0.0200 | 0.0047 | 0.0153 |
|  | V | 0.050 | 0.0430 | 0.007 |
|  | Co | 0.109 | 0.5256 | -0.0417 |
|  | Cu | 0.175 | 0.0840 | 0.091 |
|  | As | 0.01067 | 0.0335 | -0.0228 |
|  | W | 0.11 | 0.0124 | 0.0976 |
|  | Pb | 0.001 | 0.0012 | -0.0002 |
|  | Bi | 0.0004 | 0.0002 | 0.0002 |
| 0.5 | Si | 0.5093 | 0.0802 | 0.4219 |
|  | P | 0.0200 | 0.0224 | -0.0024 |
|  | V | 0.050 | 0.0470 | 0.003 |
|  | Co | 0.109 | 0.1497 | -0.0407 |
|  | Cu | 0.175 | 0.1093 | 0.0657 |
|  | As | 0.01067 | 0.0137 | -0.0030 |
|  | W | 0.11 | 0.0108 | 0.0992 |
|  | Pb | 0.001 | 0.0011 | -0.0001 |
|  | Bi | 0.0004 | 0.0006 | -0.0002 |

### 3.3 Copper Base Alloy Sample

Using GUPIXWIN software, PIXE curve obtained for copper base alloy sample is shown in the Figure 6. The matrix and trace elements identified in this sample are Cu, Ti and Sr. The major element is Cu, with concentration of 99% or greater, while the other two (Ti and Sr) are trace elements (concentration less than 1%). Standard data for Ti and Sr were unknown. The concentrations found for trace elements (Ti and Sr in copper alloy sample for different beam charges at different dates and have small standard deviation A PAR-file is created for copper base alloy sample which include all information about it. The Table 4 show the concentration of trace elements identified in copper alloy sample (more than 99% concentration) at different dates and the results obtained are found to be in close agreement. The standard deviation for copper base alloy sample is given in Table 4 and plotted in plotted in Figure 7. It is clear from the Figure that the difference of measured and already known data is negligibly small.

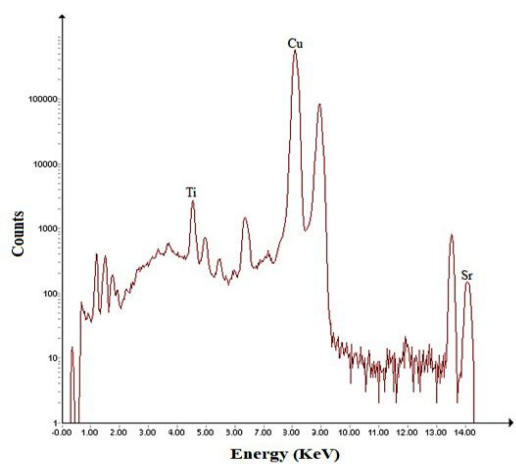

Figure 6: PIXE spectrum for copper base alloy sample.

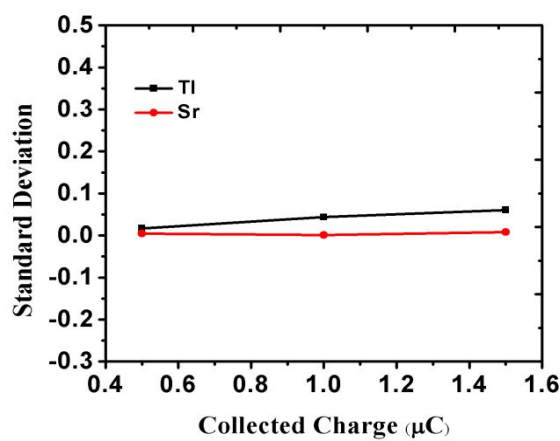

Figure 7: Dispersion plot of copper base alloy sample.

### 4. CONCLUSIONS

The main objective of this research work is to standardize PIXE technique for trace element analysis. Three samples (two NIST and one local) were analyzed using PIXE technique. Trace elements composition of measured data and standard data (known composition) were in close agreement. In NIST654B titanium alloy sample, the maximum standard deviation (σ) for Si is 0.0234 %, Cr 0.0166 %, Fe 0.0167 %, Ni 0.0078 %, Cu 0.047

% and for Sn is 0.0061 %. Similarly, in copper base alloy sample, the maximum deviation for titanium is 0.0602 % and for strontium is 0.0079 %. The standard deviation values for all trace elements in NIST654B titanium alloy and copper base alloy sample were very small (less than 1 %). In NIST1155 iron alloy sample, the maximum difference is very small (less than 1 %) for each trace element. The very small (less than 1 %) differences in elemental composition of measured and known values show that PIXE is versatile, feasible and reliable technique for trace elements analysis in thick samples.

Table 4: Comparison of measured and actual composition of copper base alloy sample for different beam of charges at different dates.

| Charge (μC) | Element | $b_1$% | $b_2$% | $b_3$% | <b>% | Standard deviation (σ)% |
|---|---|---|---|---|---|---|
| 1.5 | Ti | 0.22 | 0.08 | 0.08 | 0.12 | ±0.0602 |
| | Sr | 0.07 | 0.08 | 0.0614 | 0.07 | ±0.0079 |
| 1 | Ti | 0.16 | 0.07 | 0.07 | 0.10 | ±0.0435 |
| | Sr | 0.0795 | 0.0989 | 0.0754 | 0.08 | ±0.0012 |
| 0.5 | Ti | 0.11 | 0.07 | 0.0832 | 0.08 | ±0.0168 |
| | Sr | 0.0860 | 0.0914 | 0.09 | 0.0905 | ±0.0048 |